\documentclass[english]{emulateapj}
\usepackage{amsmath}
\usepackage{color}

\makeatletter

\@ifundefined{textcolor}{}
{%
 \definecolor{BLACK}{gray}{0}
 \definecolor{WHITE}{gray}{1}
 \definecolor{RED}{rgb}{1,0,0}
 \definecolor{GREEN}{rgb}{0,1,0}
 \definecolor{BLUE}{rgb}{0,0,1}
 \definecolor{CYAN}{cmyk}{1,0,0,0}
 \definecolor{MAGENTA}{cmyk}{0,1,0,0}
 \definecolor{YELLOW}{cmyk}{0,0,1,0}
 }

\makeatother

\usepackage[normalem]{ulem}  

\usepackage{babel}

\usepackage{natbib}

\shorttitle{Burst Oscillation Models}
\shortauthors{Mahmoodifar and Strohmayer}

\begin{document}

\title{X-ray Burst Oscillations: From Flame Spreading to the Cooling Wake}
\author{Simin Mahmoodifar and Tod Strohmayer\\ {\normalfont 
Astrophysics Science Division and Joint Space-Science Institute, NASA's 
Goddard Space Flight Center, Greenbelt, MD 20771, USA} } 

\begin{abstract}
Type I X-ray bursts are thermonuclear flashes observed from the
surfaces of accreting neutron stars (NSs) in Low Mass X-ray
Binaries. Oscillations have been observed during the rise and/or decay
of some of these X-ray bursts. Those seen during the rise can be well
explained by a spreading hot spot model, but large amplitude
  oscillations in the decay phase remain mysterious because of the
  absence of a clear-cut source of asymmetry. To date there have not
been any quantitative studies that consistently track the oscillation
amplitude both during the rise and decay (cooling tail) of
bursts. Here we compute the light curves and amplitudes of
oscillations in X-ray burst models that realistically account for both
flame spreading and subsequent cooling. We present results for
several such ``cooling wake'' models, a
``canonical'' cooling model where each patch on the NS
surface heats and cools identically, or with a latitude-dependent
cooling timescale set by the local effective gravity, and an
``asymmetric'' model where parts of the star cool at
significantly different rates. We show that while the
canonical cooling models can generate oscillations in the tails of
bursts, they cannot easily produce the highest observed modulation
amplitudes. Alternatively, a simple phenomenological model with
asymmetric cooling can achieve higher amplitudes consistent with the
observations. 

\end{abstract}
\section{Introduction}

\begin{figure}
\includegraphics[width=3.30in, height=2.8in]{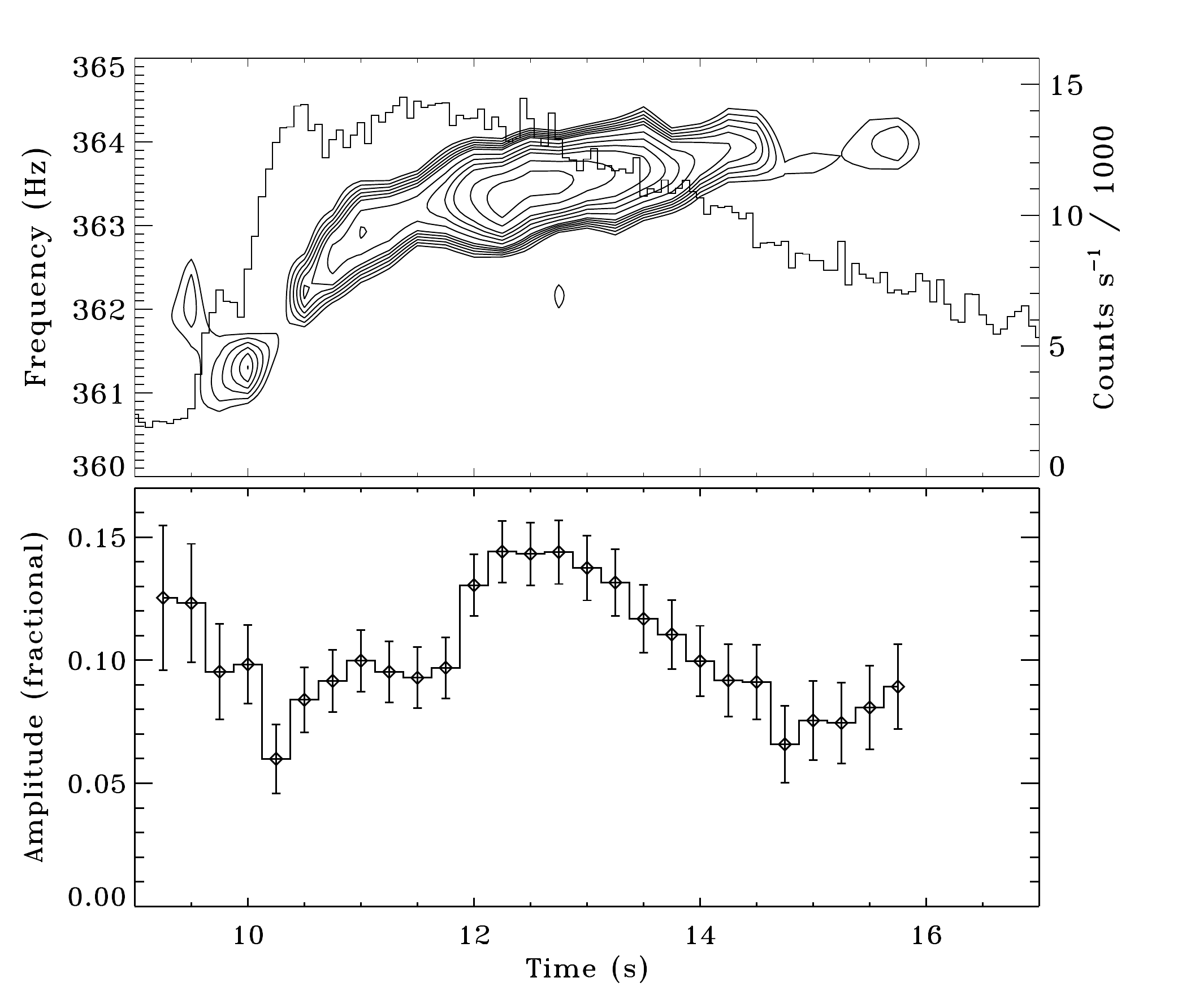}
\caption{\label{fig:burst1728} Dynamic power spectrum overplotted on
  the PCA light curve (top) for the 1997 September 20 burst from 4U
  1728-34, and the fractional amplitude (half-amplitude, see Section
    3) of oscillations during the burst (bottom). The contour levels that are plotted on the top panel are [20, 24, 28, 32, 36, 40, 50, 60, 70, 80, 100, 120]. Power spectra
    (and amplitudes) were computed from overlapping 1 s intervals,
    with a new interval starting every 1/4 s. The light curve has
    1/16 s time bins.}
\end{figure}

Type I X-ray bursts are thermonuclear explosions observed from neutron
stars (NSs) in many Low Mass X-ray Binaries (LMXBs). In these systems
the NS accretes H- and/or He-rich matter from its companion star and
when enough fuel is accumulated it can burn unstably causing periodic
thermonuclear flashes on the NS surface
\citep{2006csxs.book..113S,2008ApJS..179..360G,1976Natur.263..101W,
  1978ApJ...220..291L}.  During a burst the X-ray flux rises by a
factor of $\sim 10-20$ in a couple of seconds and then decays on a
longer timescale as the surface cools. Oscillations have been detected
during the rise and/or decay of some of these X-ray bursts with
frequencies that are within a few Hz of the NS spin frequency
\citep{1996ApJ...469L...9S, 2012ARA&A..50..609W}.  These ``burst
oscillations'' are rotationally induced modulations resulting from an
asymmetric temperature distribution on the NS surface. Their
properties during the rise can be explained by a spreading hot spot
that engulfs the star in about 1 s 
  \citep{1997ApJ...487L..77S,2014ApJ...792....4C}, however,
oscillations observed during the decay phase (tail) of bursts have
been harder to understand, as the source of asymmetry is less clear
cut.  Figure \ref{fig:burst1728} shows an example of an X-ray burst
observed with {\it the Rossi X-ray Timing Explorer (RXTE)} from the LMXB 4U
1728-34 on 1997 September 20, which shows oscillations both during
burst rise and decay. As shown in the lower panel, during the rise the
amplitude of the oscillations decreases until the peak of the burst,
then it slowly increases in the tail until it reaches a maximum of
$\approx 15 \%$ and then decays again. Here and throughout we use the so-called half-amplitude (see \S 3
  for a detailed explanation).
  
Several models have been proposed to explain the tail oscillations. It
was suggested by \cite{2004ApJ...600..939H} that they might be due to
surface oscillation modes, such as r-modes (see also
\cite{2005MNRAS.361..659L}), however, it remains unclear whether the
modes can be excited to the amplitudes required to account for the
observed X-ray modulation amplitudes (see Figure 1), and estimates of
their mode frequencies in the co-rotating frame are uncomfortably high
to be consistent with the observed closeness of the burst oscillation
and stellar spin frequencies
\citep{2008MNRAS.385.1029B,2002ApJ...580.1048M}.

Alternatively, \cite{2000ApJ...544..453C} suggested that the tail
oscillations might be due to a ``cooling wake,'' the temperature
asymmetry due to cooling of the NS surface since it takes a finite
time for both the atmosphere to cool and the burning to spread around
the star. \cite{2002ApJ...566.1018S} showed that
the temperature gradient will drive a zonal thermal wind moving
opposite to the NS rotation, and suggested that if an
inhomogeneous feature such as a vortex were trapped in it, this could
produce a flux modulation.

The timescale and geometry of the flame spreading are important, even
for oscillations in the tail, because they set the ``initial
conditions'' for subsequent cooling. Spitkovsky et al. (2002) first
analyzed global hydrodynamical flows on an accreting NS and estimated
the burning front speed taking into account the Coriolis effects. It
was subsequently shown by \cite{2007ApJ...666L..85B} and
\cite{2008MNRAS.383..387M} that these effects are important for
explaining observations of burst light curves and oscillation
amplitudes (see also \cite{2014ApJ...792....4C}).  Spitkovsky et
al. (2002) also argued that ignition is likely to start at the equator
where the effective gravitational acceleration is lowest on a fast
rotating NS, but later \cite{2007ApJ...657L..29C} showed that,
depending on the accretion rate, ignition could occur on or off the
equator. More recently, Cavecchi et al. in a series of papers
  \citep{2013MNRAS.434.3526C, 2015MNRAS.448..445C, 2015arXiv150902497C}
  presented results of 2D hydrodynamical simulations of Type I
  X-ray bursts, including the effect of rotation and magnetic field
  which confirmed earlier results by Spitkovsky et al. (2002) regarding
  the importance of the Coriolis force on the flame propagation and
  showed that the magnetic field can also have an influence on the
  flame propagation and ignition location that, depending on the field
  strength, may be comparable to the Coriolis force.

However, to date there have not been any detailed, quantitative studies
of the properties of tail oscillations that can be produced by
realistic ``cooling wake'' models. Here, we attempt to address
the question of whether such models can naturally account for the
relatively large amplitudes seen in some bursts, or whether additional
physical effects are needed. The plan of the paper is as follows. We
first summarize our methods for modeling burst light curves, including
ignition, spreading and subsequent cooling. We then present
  results for several models. First we explore two ``canonical''
  cooling models, where each patch on the NS surface heats and cools
  in a prescribed way.  In the first such model the heating and
  cooling functions are assumed to be identical for all parts of the
  star. In the second case we allow for latitude dependent variations
  in the cooling timescale, as might occur if the accretion rate (and
  thus column depth) were to vary with latitude. Such variations are
  expected for rapidly rotating stars due to latitudinal variations in
  the effective surface gravity (see \cite{2002ApJ...566.1018S};
  \cite{2007ApJ...657L..29C}). We show that while such models do
  produce oscillations in the cooling tail, they cannot easily
  generate the large amplitudes seen in some bursts.  Next, we explore
  a phenomenological ``asymmetric'' cooling model where parts of the
  star cool at significantly different rates. We show that this simple
  model can achieve higher amplitudes consistent with the largest
  observed. Finally, we briefly discuss several physical effects that
  could induce asymmetric cooling.
  
\section{Method} 
 
To model X-ray burst light curves we assume the burst ignites a hot
spot locally somewhere on the star and burning 
then spreads to engulf the entire surface.  We assume that surface
elements comprising the hot spot emit a black body spectrum, and we
take into account in our computations both special- and general
relativistic effects, including Doppler shifts and boosting,
aberration, gravitational redshifts and light-bending in the
Schwarzschild geometry (the so-called Schwarzschild$+$Doppler
approximation, see \cite{1998ApJ...499L..37M,2006MNRAS.373..836P}).
We also include an angle dependent emissivity in our models using an
approximation from J. Poutanen (2012, private communication), where the
spectrum is given by the Planck
function, but the angular distribution is assumed to be that produced
by a coherent electron scattering dominated, plane-parallel,
semi-infinite atmosphere.

On a fast rotating NS the burning region is unlikely to spread
uniformly over the surface and one needs to include the effect of the
Coriolis force
\citep{2002ApJ...566.1018S,2007ApJ...666L..85B,2008MNRAS.383..387M}.
Spitkovsky et al. (2002) estimated the flame speed in a NS ocean as
follows:
\begin{equation}
v_{flame} \sim [\frac{g h_{hot}}{t_n}\frac{1/t_{fr} +\eta/t_n}{f^2+(1/t_{fr} 
+ \eta/t_n)^2}]^{1/2}
\end{equation}
where $t_n$ is the timescale of the thermonuclear burning which is set
by the composition of the burning material, $t_{fr}$ is the frictional
coupling timescale between the top and bottom layers of the burning
region, $h_{hot}$ is the scale height of the ocean behind the
burning front, $f=2\Omega \cos\theta$ is the Coriolis parameter,
where $\Omega$ is the angular spin frequency of the NS and $\theta$ is
the colatitude of the spot center, and $\eta$ is a constant of order unity.

When frictional coupling is weak, (i.e. $t_{fr}\gg t_n$ and $t_{fr}
\gg 1/f$):
\begin{equation}
v_{flame}=\frac{v_{pole}}{\cos\theta}
\end{equation}
where $v_{pole}\sim \sqrt{g h_{hot}}/2\Omega t_n$. 

In the case of strong friction ($t_{fr}\lesssim t_n$ and 
$t_{fr}\ll 1/f$):
\begin{equation}
v_{flame}\sim(\frac{gh_{hot}t_{fr}}{t_n})^{1/2}
\end{equation}

The flame speed reaches its maximum value in the intermediate regime
when friction is important (i.e. $t_{fr}\lesssim t_n$) and $t_{fr}=
1/f$:
\begin{equation}
v_{flame}=\frac{v'_{pole}}{\sqrt{\cos\theta}}
\end{equation}
where $v'_{pole}\sim \sqrt{gh_{hot}/(4\Omega t_n)}$.

\cite{2013MNRAS.434.3526C} performed hydrodynamic simulations of the
lateral propagation of a deflagrating, vertically resolved flame on
the surface of a rotating, accreting NS and showed that the horizontal
propagation velocity is proportional to $\frac{1}{\cos\theta}$. In
our modeling we consider both cases of $1/\cos\theta$ and
$1/\sqrt{\cos\theta}$ for the latitude dependence of the flame
speed. We consider flame speeds such that the time to ignite the
entire stellar surface is less than a few seconds, so as to
approximately match the observed rise times of bursts.


\begin{figure*}
\begin{center}
\begin{tabular}{lr}
\includegraphics[width=3.30in, height=2.8in]{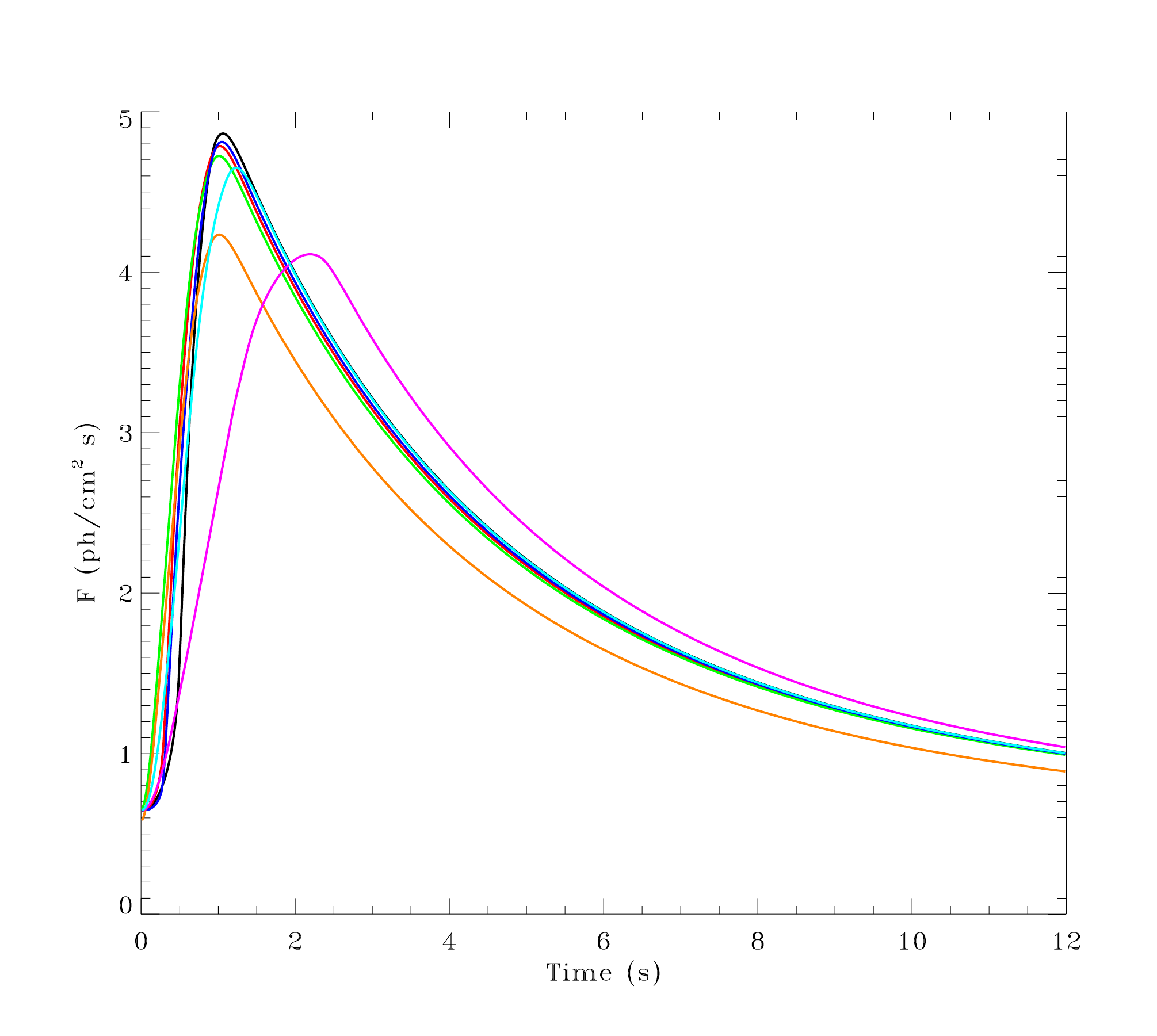}&
\includegraphics[width=3.30in, height=2.8in]{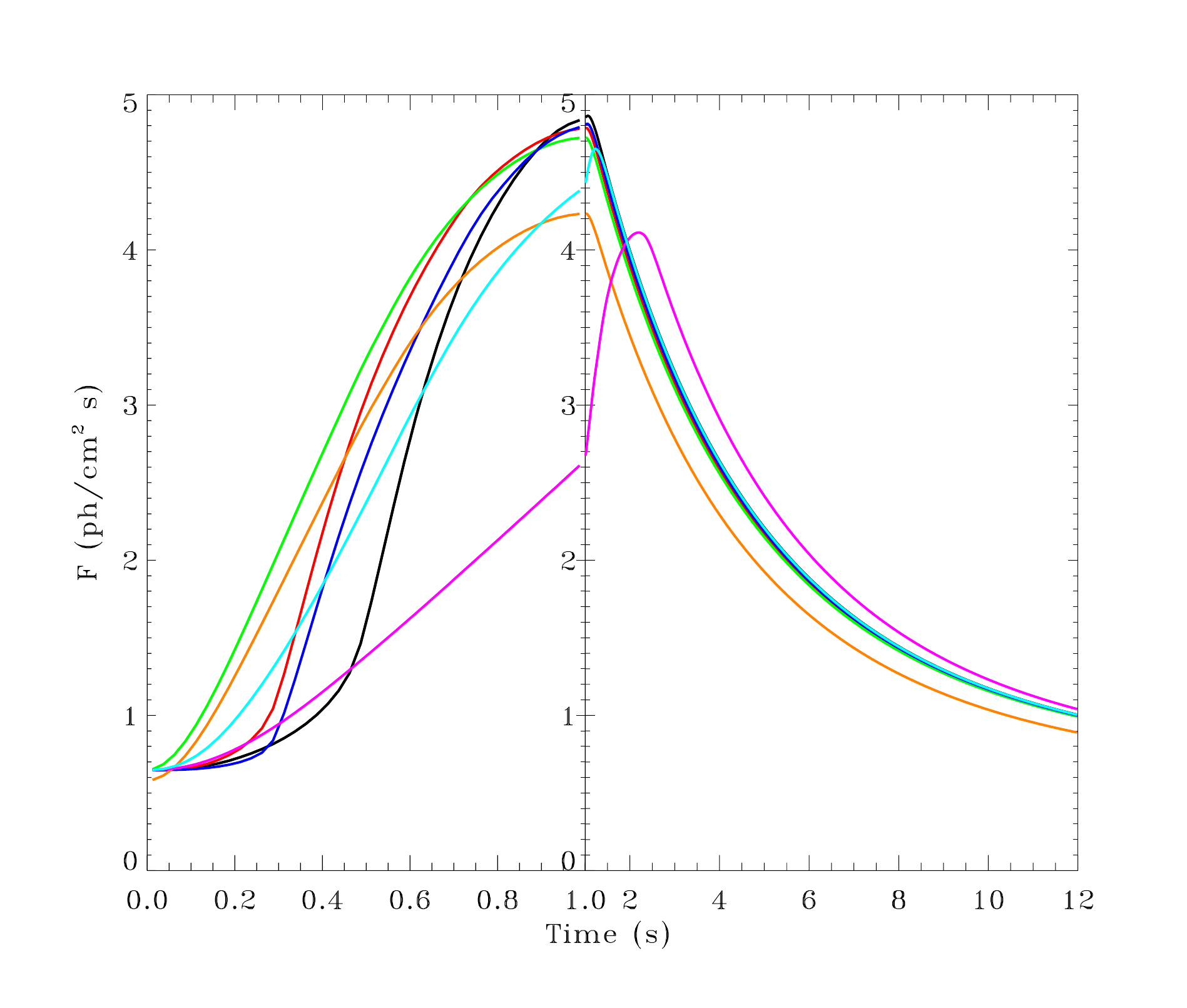} \\
\end{tabular}
\caption{\label{fig:lc_tod} Bolometric light curves for the
  ``symmetric" cooling model. The left panel shows the smoothed
  bolometric light curves during the rise and decay of the burst
  (averaged over five rotation cycles). The right panel is similar
    to the left panel, but the first second of the light curve (burst
    rise) is shown on an expanded scale. The black, red, green and
  blue curves correspond to models with $M=1.4M_{\odot}$, $R$=10 km,
  $\nu$=400 Hz, $i=70^{\circ}$, $D=10$ kpc and $v_{flame}\propto
  \frac{1}{\cos\theta}$ with $\theta_s= 10, 30, 85$ and $150^{\circ}$
  respectively.  The orange and cyan curves have similar parameters as
  the green curve ($\theta_s=85^{\circ}$) but with $M=1.8M_{\odot}$
  and $v_{flame}\propto \frac{1}{\sqrt {\cos\theta}}$
  respectively. The magenta model is the same as the cyan one but with
  a spreading speed that is half of that model. Temperature parameters
  used in these models are $T_0=1.5$ keV, $\Delta T=1.5$ keV, $\Delta
  t_r=0.46$ s, $\tau_r=0.1$ s and $\tau_d=6$ s.}
\end{center}
\end{figure*}

\begin{figure*}
\begin{center}
\begin{tabular}{lr}
\includegraphics[width=3.30in, height=2.8in]{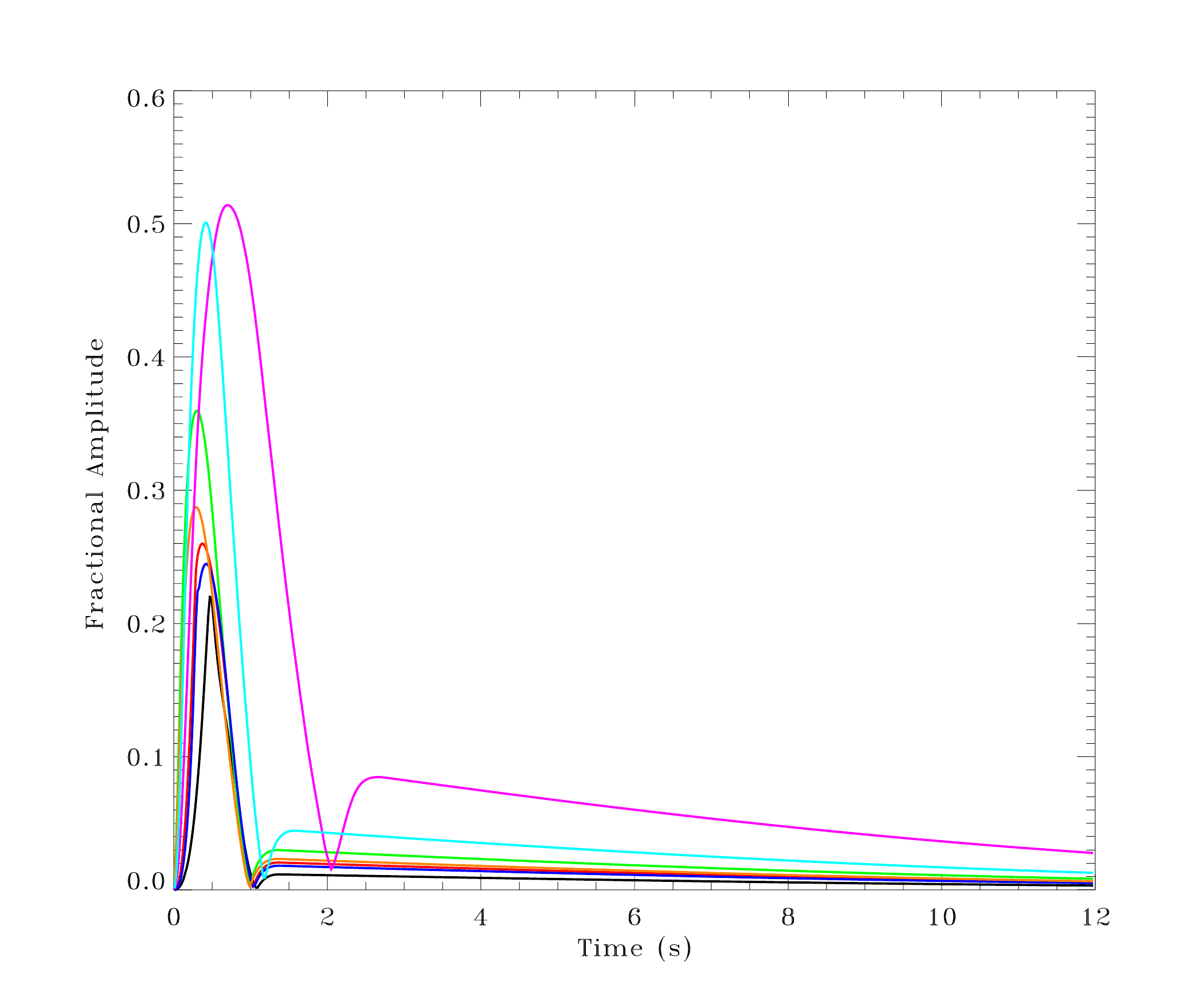}&
\includegraphics[width=3.30in, height=2.8in]{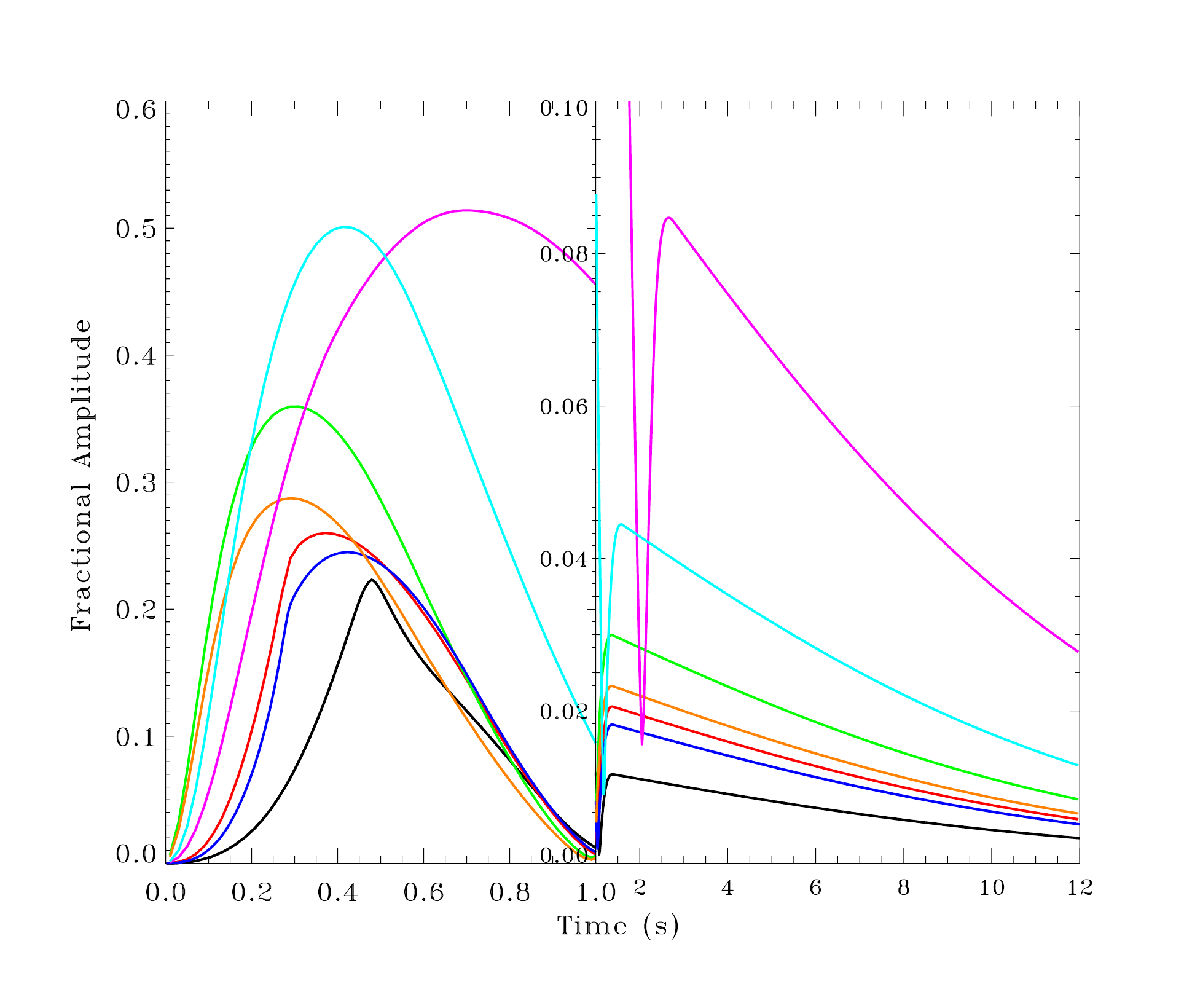} \\
\end{tabular}
\caption{\label{fig:famps_tod} Fractional oscillation amplitude
  vs. time for the ``symmetric" cooling model. The color coding is the
  same as Figure \ref{fig:lc_tod}. On the right panel the fractional
  amplitudes are shown on an expanded scale up to $t=1$ s. Note that the vertical scales are also different before and after $t=1$ s.}
\end{center}
\end{figure*}

We consider several models to explore the
properties of burst oscillations in the tail. The first is a
``canonical'' cooling model in which we assume that, once ignited, all
surface elements of the star follow the same temperature evolution
(we refer to this as the ``symmetric'' cooling model). In this
model each surface element, once ignited, evolves in temperature
according to an exponential rise and decay. We use an expression
similar to \cite{2006ApJ...636L.121B} and
\cite{2008MNRAS.383..387M} for the temperature evolution:

\begin{equation}\label{eq:temperature}
\begin{aligned}
T =  T_0,  \qquad \qquad \qquad \qquad \qquad \qquad \qquad \qquad \qquad \quad t\leq t_{ig}\\
\quad  =  (T_0+\Delta T[1-\exp(\frac{-(t-t_{ig})}{\tau_{r}})]), \quad   t_{ig}\leq t\leq t_{ig}+\Delta t_r\\
\quad = (T_0+\Delta T_m\exp(\frac{-(t-(t_{ig}+\Delta t_{r}))}{\tau_{d}})), \quad   t\geq t_{ig}+\Delta t_r
 \end{aligned}
\end{equation}

where $t_{ig}$, $\Delta t_r$, $\tau_{r}$, $\tau_{d}$ are the
  ignition time, the temperature rise time (from ignition to peak),
  the rise timescale, and the decay timescale, respectively. In
  addition, $T_0$ and $\Delta T$ are the background (minimum) temperature and 
  temperature contrast, respectively and $\Delta T_m=\Delta T[1-\exp(\frac{-\Delta
      t_r}{\tau_{r}})]$.
      
1D hydro simulations of X-ray bursts \citep{1980ApJ...241..358T,2004ApJS..151...75W,2006ApJ...639.1018W} give the light curve for a single patch on the
  star. The shape of the light curve in these models is sensitive to
  the composition of the accreted layer and the burning regime. The
  temperature profile that we use here gives a reasonable
  approximation to the shape of a single patch light curve in these
  simulations. Temperature parameters used in our models are $T_0=1.5$
  keV, $\Delta T=1.5$ keV, $\Delta t_r=0.46$ s, $\tau_r=0.1$ s and
  $\tau_d=6$ s. These parameters result in a burst rise time of about
  1 s and a decay time of about 11 s. Since there is a finite time over which the burning spreads, there
will always be some temperature gradient across the NS surface. This
temperature asymmetry combined with rotation of the star results in
oscillations during the burst tail.

\begin{figure}
\includegraphics[width=3.30in, height=2.8in]{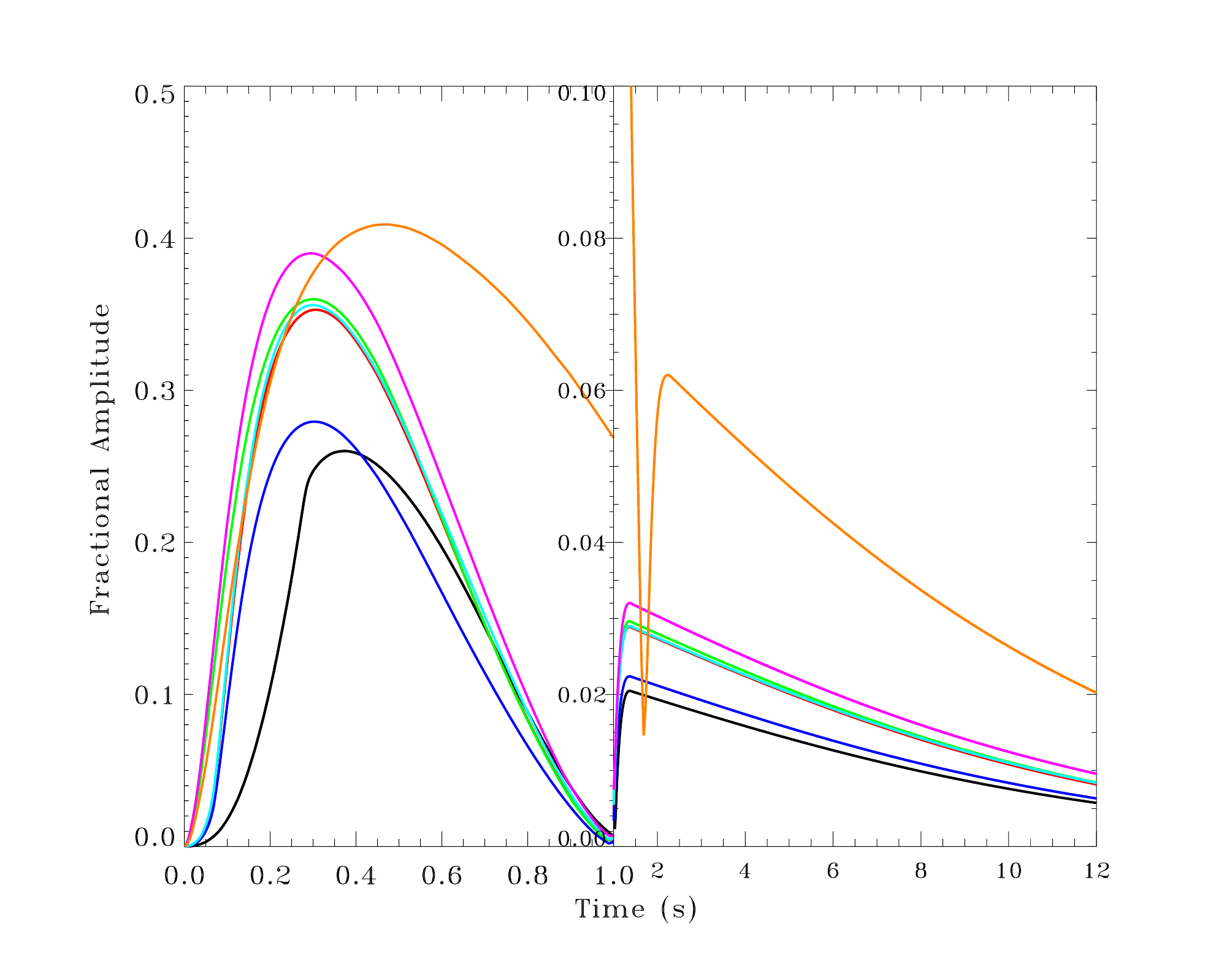}
\caption{\label{fig:famps_tod_asym} Fractional amplitudes for the
    ``canonical'' cooling model with latitude-dependent cooling
    timescales. The black, red, and green curves correspond to models
    with $M=1.4M_{\odot}$, $R$=10 km, $\nu$=400 Hz, $i=70^{\circ}$,
    $D=10$ kpc and $v_{flame}\propto \frac{1}{\cos\theta}$ with
    $\theta_s= 30, 60$, and $85^{\circ}$ respectively. The blue and
    cyan curves are the same as the red one but with $M=1.8M_{\odot}$
    and $\nu$=600 Hz, respectively (note that red, cyan and green curves almost overlap). The magenta curve has
    $M=1.4M_{\odot}$, $\theta_s=85^{\circ}$, $\nu$=600 Hz and $R$=12
    km. The orange curve has similar parameters as the magenta one,
    but with a spreading speed that is half of that model. Note that the amplitudes are shown on an expanded scale up to $t=1$ s. Also the vertical scale is different before and after $t=1$ s on this plot.}
\end{figure} 

\begin{figure*}
\begin{center}
\begin{tabular}{lr}
\includegraphics[width=3.30in, height=2.8in]{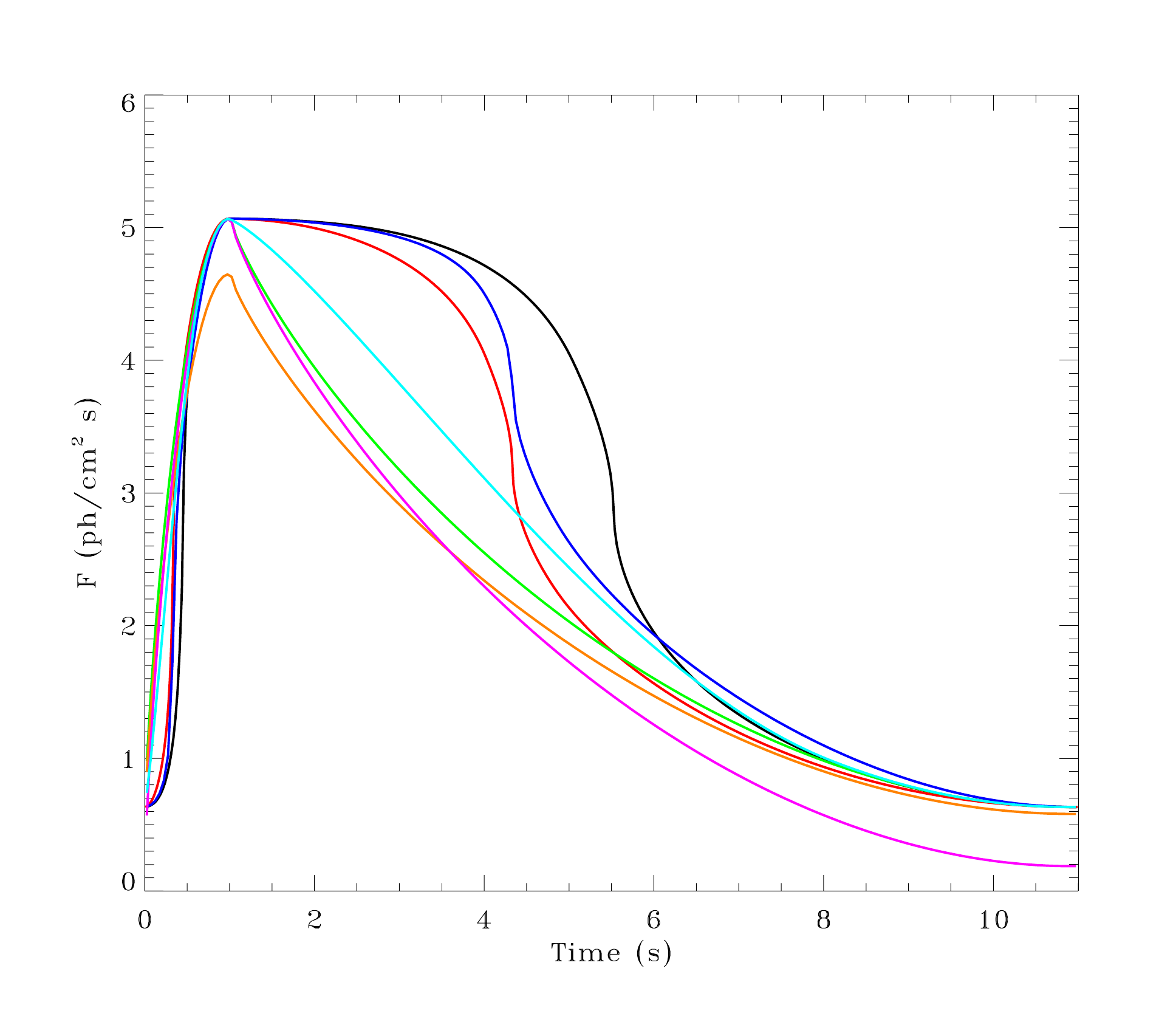}&
\includegraphics[width=3.30in, height=2.8in]{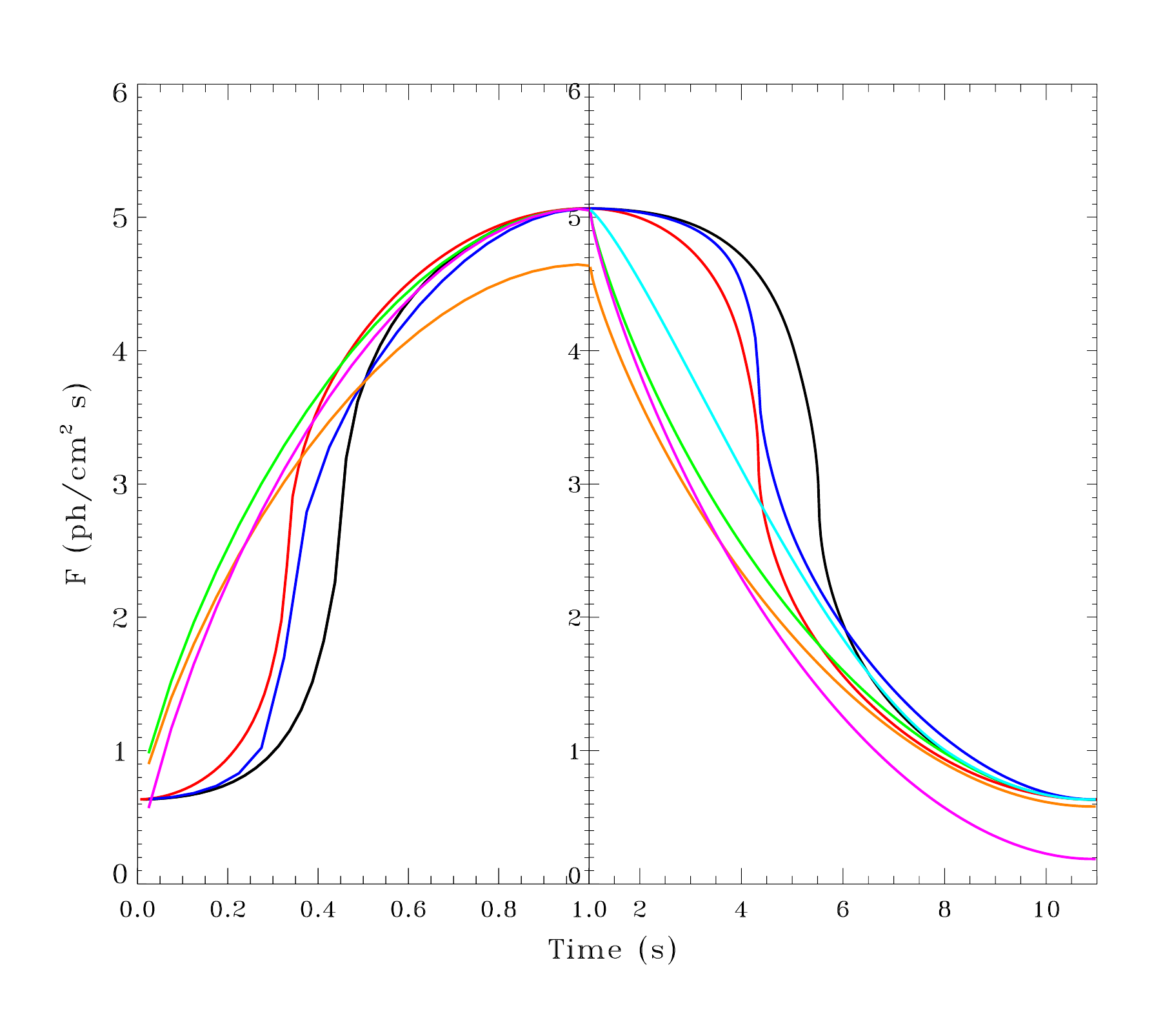} \\
\end{tabular}
\caption{\label{fig:lc_simin} Left panel shows the smoothed bolometric
  light curves for the rise and
  decay of the burst in our
  ``asymmetric'' cooling model. The black, red,
  green and blue curves correspond to models with $M=1.4M_{\odot}$,
  $R$=10 km, $\nu$=400 Hz, $i=70^{\circ}$, $D=10$ kpc, $T_h=3$ keV, $\Delta
  T=T_h-T_c=1.5$ keV and $v_{flame}\propto \frac{1}{\cos\theta}$ with
  $\theta_s= 10, 30, 85$ and $150^{\circ}$ respectively. The orange,
  magenta and cyan curves are similar to the green curve ($\theta_s=85
  ^{\circ}$) but with $M=1.8M_{\odot}$, $\Delta T$=2 keV and
  $v_{flame}\propto \frac{1}{\sqrt {\cos\theta}}$ respectively. Right
  panel shows the light curves on an expanded scale up to $t=1$ s (burst rise).}
\end{center}
\end{figure*}

\begin{figure*}
\begin{center}
\begin{tabular}{lr}
\includegraphics[width=3.30in, height=2.8in]{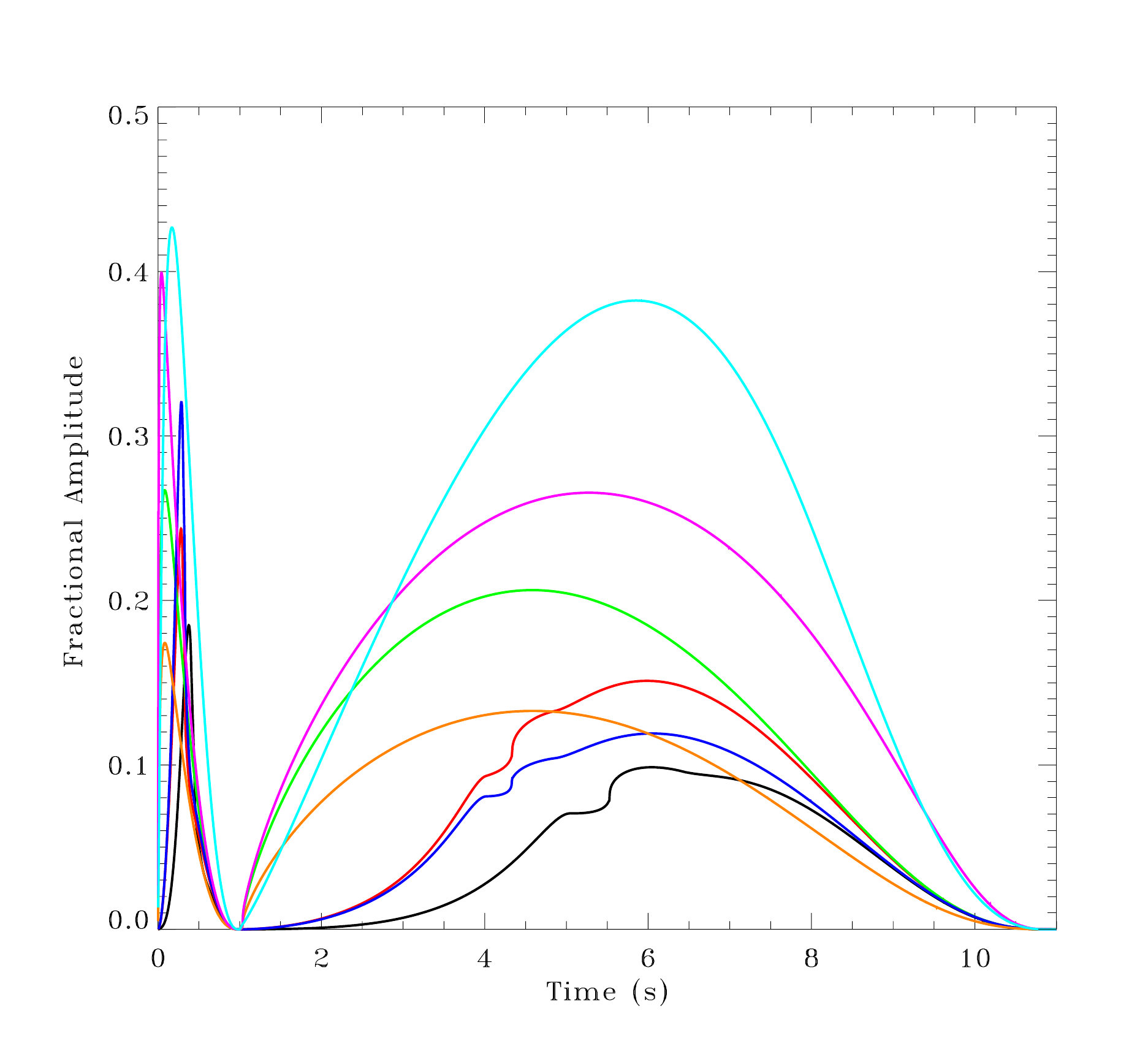}&
\includegraphics[width=3.30in, height=2.8in]{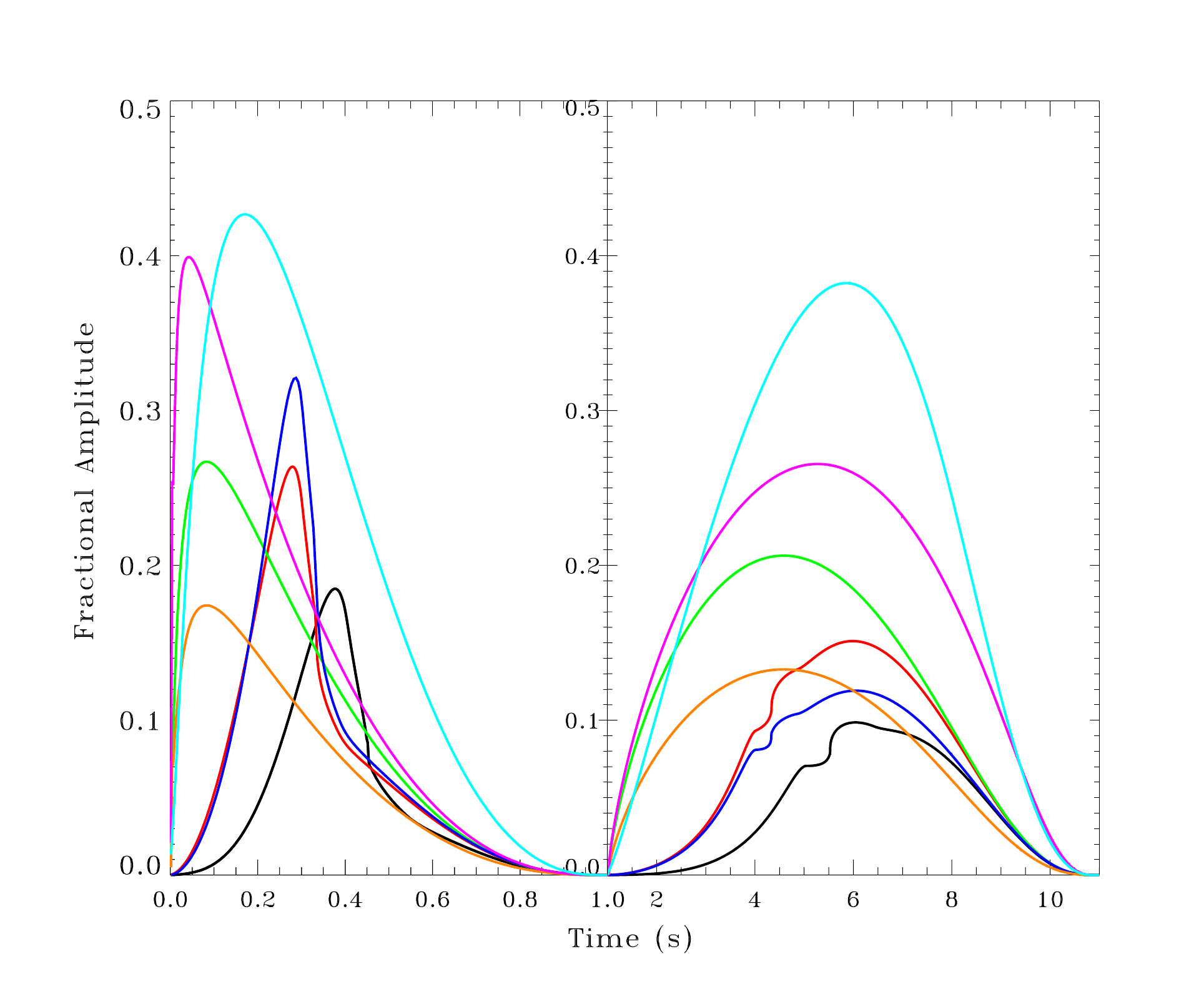} \\
\end{tabular}
\caption{\label{fig:famps_simin} Fractional oscillation amplitude
  vs. time for the ``asymmetric" cooling model. The color coding is the
  same as Figure \ref{fig:lc_simin}. In the right panel the fractional
  amplitudes are shown on an expanded scale up to $t=1$ s.}
\end{center}
\end{figure*}

  We also consider a variation
  of the ``symmetric'' model in which the
  cooling timescale, $\tau_d$, is assumed to vary with latitude
  (we call this the ``latitude-dependent'' cooling model). On a
  rapidly rotating NS the effective gravitational acceleration is
  reduced along the rotational equator and enhanced at the poles.
  This can lead to variations in the local accretion rate and hence
  the accreted column with latitude (Cooper \& Narayan 2007). Since
  the vertical cooling time is basically set by the column depth at
  which heat is released, coupled with the thermal conductivity of the
  material above, variations in the accreted column and/or composition
  across the star could in principle lead to changes in the cooling
  timescale with latitude. Since the local mass accretion rate is
  proportional to the inverse of the effective gravitational
  acceleration ($\dot{m}(\theta)\propto g_{eff}^{-1}(\theta)$)
  \citep{2002ApJ...566.1018S,2007ApJ...657L..29C}, we assume that the
  cooling (decay) timescale, $\tau_d$, in Eq.~\ref{eq:temperature}
  varies with latitude like $1/g_{eff}(\theta) \propto 1/(1+c_e
  \bar{\Omega}^2 \sin^2 \theta+ c_p \bar{\Omega}^2 \cos^2
  \theta)$. Where $c_e=-0.791+0.776 (M/R)$, $c_p=1.138-1.431 (M/R)$
  and $\bar{\Omega}=\Omega \sqrt{R^3/M}$ (in geometric units $G=c=1$;
  \citet{2014ApJ...791...78A}). Note that $g_{eff}$ is lowest at the
  equator and increases toward the pole, and therefore the mass
  accretion rate and the cooling timescale are largest at the equator
  and decrease toward the poles. While this model includes variations
  in the cooling timescale across the star, the variations retain
  symmetry about the spin axis.

Lastly, we consider a simple phenomenological model with
  ``asymmetric'' cooling. By this we mean that the variations in the
cooling timescale are no longer symmetric about the rotational
axis. In this model we do not consider the full temperature
evolution of each individual surface element. In this ``spreading
cooling wake" model we have only two temperatures, $T_h$ and $T_c$
that represent the temperature of the burning region and the rest of
the star (cool regions), respectively. In this model, when burning
spreads over the star, the temperature of each patch changes from
$T_c$ to $T_h$. The burning spreads in the same way as in the previous
model, the only difference is that at $t=t_{rise}=1$ s when the whole
surface is burning there is no temperature gradient across the
star. When cooling begins, regions that started burning earlier are
assumed to cool first and their temperature changes from $T_h$ to
$T_c$. The cooling is assumed to spread in the exact same way as the
burning but just over a longer time. Here, we
use a total duration of 10 s for
  the cooling phase, which is characteristic of observed
bursts. That is, the speed of the spreading of cooling is taken
  to be $0.1\times v_{flame}$, which gives a total decay time of 10 s.
  In this model at $t=$1 s all patches on the surface are ignited and
  have a temperature of $T_h$, but since the cooling duration is
  longer, and also the cooling is assumed to spread in a similar way
  as burning, the patches that burned first will cool first and also in
  a shorter time than other parts of the star. For example, at
  $t=$1.5 s a small region on the surface that has been ignited first
  will be already cool (with $T=T_c$), but it will take longer (up to
  10 s) for the patches that burned last to cool. Therefore the total
  time that a given patch is hot is different from any other
  patch. For example, the total time that the first burned patch is
  hot is about 1 s, the patch that burned at $t=$0.5 s will be hot
  for $\sim 5.5$ s and the last patch that burned at $t=$1 s will be hot
  for $\sim 10$ s.  This simple model approximates what may happen if
the timescale of cooling varies across the surface in a manner
  that breaks the symmetry about the rotation axis, i.e. if the heat
transport varies with position on the star. For example, if there is
significant transverse heat flow between hot and cold regions then the
timescale of cooling in the area where burning started first might be
shorter since it is initially surrounded by cooler regions, but those
that burn last are already surrounded by hot areas and therefore it
might take longer for them to cool. While it is likely that
  lateral conductive heat transport is negligible given the large
  difference between the vertical and horizontal scales in the burning
  front (Cavecchi et al. 2013), convective and hydrodynamic flows not
  fully captured by 2D hydrodynamic simulations could perhaps
  increase the transverse heat flow. Other possibilities for inducing
asymmetric cooling would be different fuel depths or composition
across the star and also varying amounts of heat flux from deeper
layers for different parts of the NS surface. 

\begin{figure*}
\begin{center}
\begin{tabular}{lr}
\includegraphics[width=3.30in, height=2.8in]{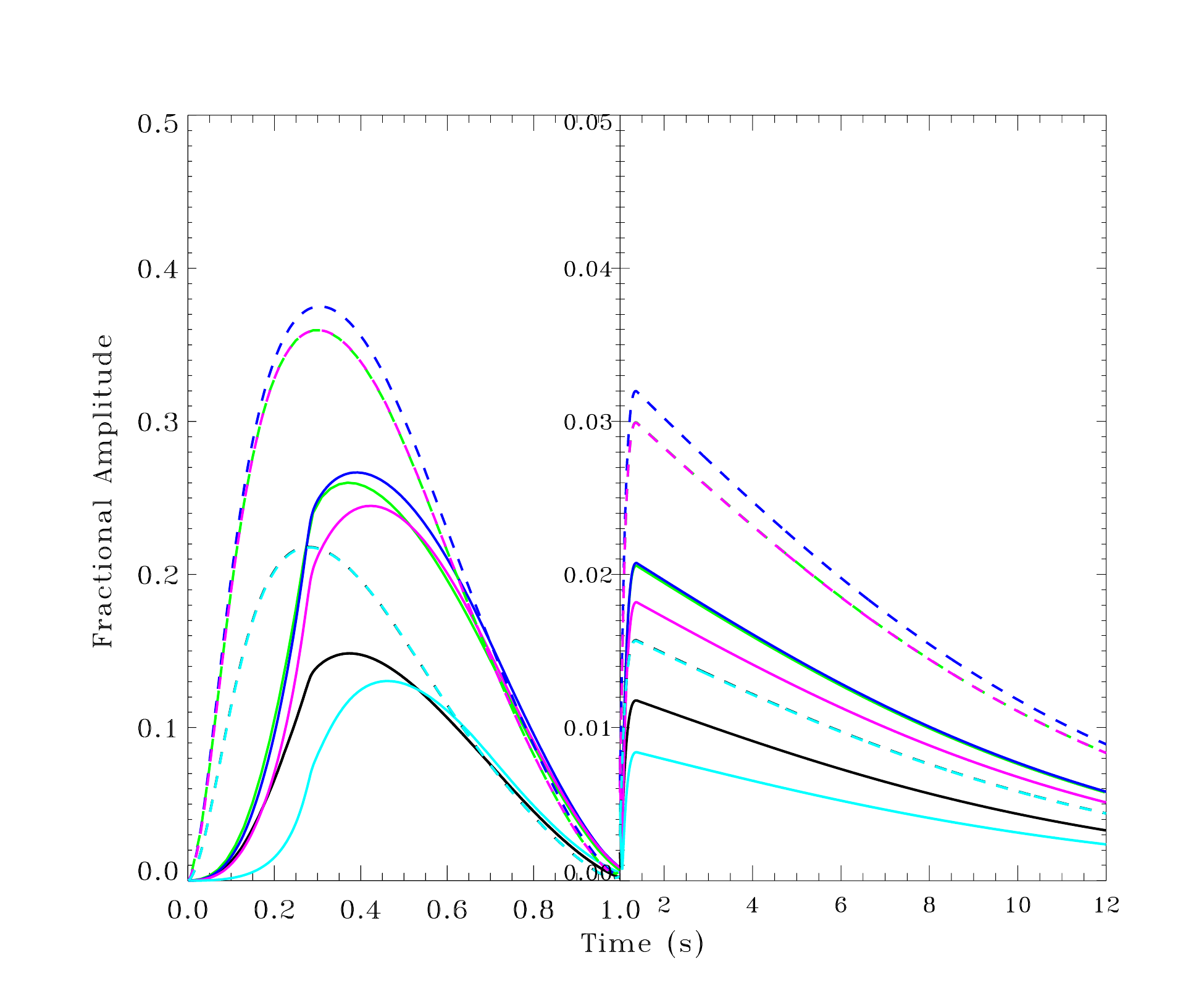}&
\includegraphics[width=3.30in, height=2.8in]{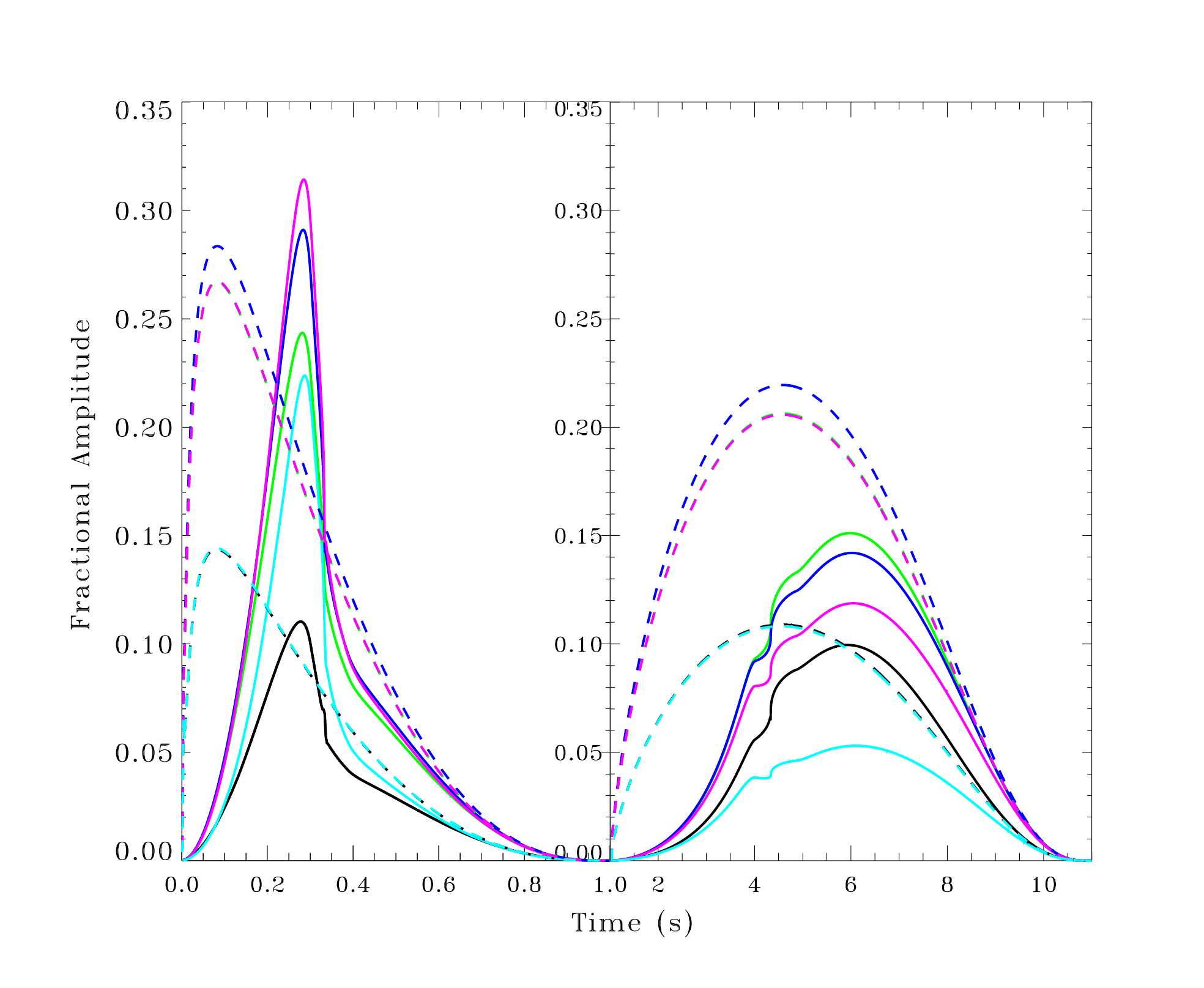} \\
\end{tabular}
\caption{\label{fig:famps_inclination} Fractional amplitude
    vs. time for different inclination angles for the ``symmetric''
    cooling model (left panel) and ``asymmetric'' model (right
    panel). In both plots the solid curves correspond to an ignition
    latitude of $\theta_s=30^{\circ}$ and the dashed curves correspond
    to an ignition latitude of $\theta_s=85^{\circ}$. The black,
    green, blue, magenta and cyan curves correspond to $i=30, 70, 90,
    110$ and $130^{\circ}$, respectively. Other parameters are the
    same as those used for the red and green models in Figures 2 and
    4. Note that the amplitudes are shown on an expanded scale up to $t=1$ s. Also the vertical scale is different before and after t=1 s on the left panel.}
\end{center}
\end{figure*}

\section{Results}

Figure \ref{fig:lc_tod} shows several burst light curves computed with
the canonical (``symmetric'') cooling model. Different curves
correspond to different ignition latitudes, $\theta_s$, NS compactness
and flame spreading speeds ($v_{flame}\propto \frac{1}{\cos\theta}$
and $\frac{1}{\sqrt{\cos\theta}}$). The black, red, green and blue
curves correspond to models with $M=1.4M_{\odot}$, $R=10$ km,
$\nu=400$ Hz, $i=70^{\circ}$, $D=10$ kpc and $v_{flame} \propto
\frac{1}{\cos\theta}$ with $\theta_s= 10, 30, 85$ and $150^{\circ}$,
respectively.  Here, $M$, $R$, $\nu$, $i$ and $D$ are the stellar
mass, radius, spin frequency, observer's inclination angle, and the
distance to the source, respectively. The orange and cyan curves have
similar parameters as the green curve (all have $\theta_s=85^{\circ}$)
but with $M=1.8M_{\odot}$ and $v_{flame}\propto \frac{1}{\sqrt
  {\cos\theta}}$, respectively, to show the effect of compactness and
flame spreading velocity on the light curve. The magenta curve has the
same parameters as the cyan curve but with a spreading speed that is
half of the cyan one.

To highlight the differences in the rising phase more clearly, in
Figure \ref{fig:lc_tod} (right panel) we also show the light curves 
on an expanded scale up to $t=1$ s. As
burning spreads over the surface the flux increases until it engulfs
the whole star, after that, since the temperature of each surface
element decays exponentially with time, the flux decreases. The shape
of the light curve during the rising phase depends on several factors
such as the ignition latitude, the flame spreading geometry and speed,
and its latitudinal dependence \citep{2008MNRAS.383..387M}.  When
ignition starts near a pole (black, red and blue curves with
$\theta_s=10, 30$ and $150^{\circ}$) the rise light curve is more
concave (rises more slowly) compared to when it starts near the
equator ($\theta_s=85^{\circ}$, see \cite{2008MNRAS.383..387M}).

Figure \ref{fig:famps_tod} shows the evolution of burst oscillation
amplitudes during the burst rise and decay for the ``symmetric''
model. The color coding is similar to Figure \ref{fig:lc_tod}. To compute
the fractional amplitudes we calculate the so-called
  half-amplitude. The light curve is phase-folded in the time
intervals of interest, and the amplitude is determined as $f =
(I_{max} - I_{min})/(I_{max} + I_{min})$, where $I_{max}$ and
$I_{min}$ are the maximum and minimum values in the phase-folded light
curves, respectively. To the extent that the light curves are
  consistent with sinusoids, this is equivalent to fitting the
  phase-folded light curve with the model $A+B \sin (2 \pi \nu
  t+\phi_0)$, and defining the fractional amplitude as $B/A$. This
  definition of fractional amplitude is $\sqrt{2}$ times larger than
  the rms amplitude used in many papers. Therefore, to compare the
  rms amplitudes with our amplitudes, they should be multiplied by
  $\sqrt{2}$.

The maximum of the tail amplitudes in the
  ``symmetric'' model ranges from ~1.5\% to 8\% depending on the ignition
latitude, NS compactness and the flame spreading geometry/speed. The
temperature asymmetry combined with rotation of the star produces
oscillations during the decaying phase, but since the temperature
gradient across the surface is not very large after burning has spread
over the whole surface, the amplitude does not get large enough to
explain the observations where the amplitudes in the tail can be $\geq
15 \%$ (see Figure 1; 
  \cite{2008ApJS..179..360G,2002ApJ...581..550M,2001ApJ...551..907V}). Note
that the amplitudes only slightly increase by increasing $\Delta T$ in
Eq.~\ref{eq:temperature}.  The largest tail amplitude is produced
  by the model with the longest (slowest) rise-time (spreading
  speed). This is not too surprising since there is a longer time for
  cooling to introduce a temperature gradient when the spreading speed
  is slower (assuming a constant cooling timescale).  However, some
  bursts with large tail amplitudes have fast ($\approx 1$ s) rises
  (see Figure 1), so this effect cannot account for all large observed
  tail amplitudes.

In Figure \ref{fig:famps_tod_asym} we show the evolution of the
  fractional amplitudes for several models with latitude-dependent
  cooling timescales.  The black, red, and green curves correspond to
  models with $M=1.4M_{\odot}$, $R$=10 km, $\nu$=400 Hz,
  $i=70^{\circ}$, $D=10$ kpc, $v_{flame}\propto \frac{1}{\cos\theta}$
  and $\theta_s= 30, 60$, and $85^{\circ}$, respectively. The blue and
  cyan curves are the same as the red one but with $M=1.8M_{\odot}$
  and $\nu$=600 Hz, respectively. The magenta curve has
  $M=1.4M_{\odot}$, $\theta_s=85^{\circ}$, $\nu$=600 Hz and $R$=12 km,
  and the orange model has similar parameters to the magenta one, but
  with a spreading speed that is half of that model. Increasing both
  the spin frequency and stellar radius increases the variations in
  effective gravitational acceleration from equator to pole, thus
  making the difference between cooling timescales at the equator and
  pole larger. This causes a larger temperature asymmetry across the
  star and leads to somewhat larger fractional amplitudes. Comparing the red
  and cyan curves, which almost overlap, and also the green and
  magenta curves we note that increasing the frequency has a smaller
  effect on increasing the amplitudes than the radius. However, comparing
  Figure \ref{fig:famps_tod_asym} with Figure \ref{fig:famps_tod} we see
  that the fractional amplitudes in the burst tail only slightly
  increase with the inclusion of a latitude-dependent cooling
  timescale, thus, this model is still not able to explain the
  largest observed amplitudes in burst tails.

Figure \ref{fig:lc_simin} shows the light curves for the
phenomenological model with ``asymmetric'' cooling. The black, red,
green and blue curves correspond to models with $M=1.4M_{\odot}$,
$R$=10 km, $\nu$=400 Hz, $i=70^{\circ}$, $D=10$ kpc, $T_h=3$ keV,
$\Delta T=T_h-T_c=1.5$ keV and $v_{flame}\propto \frac{1}{\cos\theta}$
with $\theta_s= 10, 30, 85$ and $150^{\circ}$ respectively. The
orange, magenta and cyan curves are similar to the green curve
($\theta_s=85^{\circ}$) but with $M=1.8M_{\odot}$, $\Delta T$=2 keV
and $v_{flame}\propto \frac{1}{\sqrt {\cos\theta}}$, respectively. As
for the ``symmetric'' model, one can see a clear difference in the
shape of the light curves during the rising phase depending on whether
the ignition starts near the pole or the equator
\citep{2008MNRAS.383..387M}. Also, for models where ignition starts
near the pole the light curve decays more slowly in the first few
seconds after the peak compared with near-equatorial ignition. This is
due to our assumption for this model that cooling spreads in the same
way as burning (i.e., with the same angular dependence as from the
Coriolis effects), in which case it would take longer for the cooling
to reach the equator.


Figure \ref{fig:famps_simin} shows the fractional amplitudes
during the burst rise and decay for the ``asymmetric'' cooling
model. The color coding is similar to Figure \ref{fig:lc_simin}.
During the burst rise as the hot spot gets larger, the fractional
amplitude increases, until burning reaches the equatorial region and
quickly wraps around the equator due to the Coriolis effect. This then
results in more symmetric burning and therefore smaller fractional
amplitudes. As the burning spreads over a larger portion of the star
and also the total flux increases the fractional amplitude continues
to decrease. When burning engulfs the whole star there is by
assumption no longer any temperature asymmetry in this model and the
fractional amplitude goes to zero.\footnote{This is due to the
    simplification in the asymmetric model where the temperature
    changes like a step function as opposed to the gradual temperature
    change in the ``canonical'' models where there is always some
    temperature difference between different regions, even in the peak
    of the burst. If there was a gradual change from $T_c$ to $T_h$
    and vice versa, the amplitude would not go to zero at the peak.}
After that, as the star begins to cool, the temperature asymmetry and
therefore the fractional oscillation amplitude grows. The amplitude of
the tail oscillations can be significantly larger in this model
compared to the ``canonical'' models, and ranges from less than 10\%
to about 35\% depending on the compactness of the star, the
temperature gradient and the flame spreading speed in this
model. The sudden rise in amplitudes around $t=4-6$~s in the red,
  blue and black curves is due to the rapid cooling near the equator,
  and therefore a sudden decrease in the total flux. This is because
  when the cooling front reaches the equator it spreads around the
  star and decreases the total flux, but since the spreading is more
  or less symmetric around the equator it doesn't cause a big
  temperature asymmetry, therefore the amplitude of the oscillations
  stays approximately the same, but since the total flux decreases,
  the fractional amplitude increases at that point. In general the
amplitude of the tail oscillations are larger when $v_{flame}\propto
\frac{1}{\sqrt{\cos\theta}}$ compared to $v_{flame}\propto
\frac{1}{\cos\theta}$, assuming that the burning engulfs the star in
$\sim 1$ s in both cases. The reason is that in the former case
it takes longer for the burning to spread around the equator and
therefore there will be more temperature asymmetry compared to the
latter. We also note that as the compactness of the star
increases the oscillation amplitudes decrease, since at each time a
larger portion of the surface would be visible.

Since we are interested in exploring the largest amplitudes that
  can be produced in the cooling tails of bursts we used a favorable
  observer's inclination angle of $70^{\circ}$ for the results shown
  in Figures 2 - 6.  However, for completeness, in
  Figure \ref{fig:famps_inclination} we also explored the effect of
  inclination angle on the amplitude of burst oscillations. The left
  panel in this figure corresponds to the ``symmetric'' cooling model,
  and the right panel to the ``asymmetric'' model. In both plots the
  solid and dashed curves correspond to an ignition latitude of
  $\theta_s=30$ and $85^{\circ}$, respectively. The black, green,
  blue, magenta and cyan curves represent $i=30, 70, 90, 110$ and
  $130^{\circ}$, respectively, with $M=1.4M_{\odot}$, $R$=10 km,
  $\nu$=400 Hz, $D=10$ kpc, $v_{flame}\propto \frac{1}{\cos\theta}$
  and $\Delta T=1.5$keV.  In the ``symmetric'' model the oscillation
  amplitudes increase as the inclination angle gets closer to
  $90^{\circ}$, where at each time more temperature asymmetry would be
  observable. In the cases where the ignition starts near the equator
  there is a symmetry between inclination angles above and below the
  equator. For example, for $\theta_s=85^{\circ}$, the amplitudes are
  almost the same for $i=70^{\circ}$ and $110^{\circ}$ and for $i=30^{\circ}$ and
  $130^{\circ}$. We note that such symmetry doesn't exist when the
  ignition starts off the equator. In the ``asymmetric'' model the
  effect of the inclination angle on oscillation amplitudes are
  similar to the ``symmetric'' model for the cases where the ignition
  starts near the equator, but for off-equatorial ignitions,
  $i=90^{\circ}$ no longer corresponds to the highest amplitudes. 

\section{Summary and Conclusions}

In this paper we have computed for the first time the light curves and
amplitudes of oscillations in X-ray burst models that realistically
account for both flame spreading and subsequent cooling. This allows
us to consistently track the oscillation amplitude both during the
rise and decay (tail) in these models.  We have compared results
  from several models. We consider two variations of ``canonical''
  cooling models, one ``symmetric'' model in which each patch on the
  NS heats and cools in the same manner, and another which allows for
  variations in the cooling timescale with latitude, as could result
  from rotationally induced changes in the effective gravity. We also
  explored a simple ``asymmetric'' model where parts of the star that
ignite first cool faster than those that burn last. This may happen,
for example, if there is significant transverse heat exchange between
hot and cold regions, or if the local cooling time is shorter near the
ignition site than elsewhere on the star. We showed
that the ``canonical'' cooling models can generate oscillations in the
tails of bursts, but they cannot easily produce the highest observed
modulation amplitudes. One way to increase the modulation amplitudes
in such models is to increase the temperature contrast during the
burst.  For a fixed cooling timescale this can be done by decreasing
the flame speed (increasing the rise time), but to reach substantially
higher amplitudes one requires rise times longer than typically
observed. Nevertheless, a signature of such a ``canonical'' cooling
wake would be a positive correlation between burst rise time and tail
oscillation amplitude.  Another way to increase the modulation
  amplitudes in the ``canonical'' model is to include a
  latitude-dependent cooling timescale in the temperature
  evolution. We showed in Figure \ref{fig:famps_tod_asym} that although
  this will boost the amplitudes, it is still not large enough to
  explain the observed amplitudes in some of the burst tails.

 On the other hand, a relatively simple phenomenological model with
 asymmetric cooling, where the speed of the cooling wake is
   different in different regions on the star, and is not symmetric
   about the rotation axis, can achieve higher amplitudes consistent
 with the highest observed.  While the particular asymmetric model we
 employ is not rigorously self-consistent, and it may tend to
 overestimate actual cooling asymmetries, its results demonstrate that
 asymmetric cooling processes can boost the amplitude of tail
 oscillations, and are thus important areas for further
 research. Indeed, the amplitude evolutions of several of the models
 in Figure \ref{fig:famps_simin} have some qualitative
 similarities to that of the 4U 1728-34 burst shown in Figure 1, in
 particular the rise in amplitude several seconds after the peak.  

A number of physical effects could plausibly lead to asymmetric
cooling, and we briefly discuss several here. As we noted previously,
significant transverse heat flow could lead to such asymmetries. The
ignition site is initially surrounded by cooler regions, thus, heat
flow laterally away from the hot spot could cool it faster than
regions which are last to burn, since they are surrounded by hot,
already ignited fuel. Previous work has shown that the vertical scale
height within the front is much smaller than the relevant transverse
length scales (for example, the Rossby adjustment radius, Spitkovsky
et al. 2002; Cavecchi et al. 2013). This suggests that thermal
conduction is unlikely to result in significant transverse heat
flow. However, 2D simulations indicate that larger scale horizontal
(zonal) and convective flows can be set-up by the burning.  These
flows could perhaps transport heat in a manner which induces some
level of asymmetric cooling like that present in our simple
``asymmetric'' model.  Fully 3D hydrodynamic simulations will likely
be needed to further explore the likely magnitude of any such effects,
and they can then be incorporated into fully self consistent light
curve models.

\acknowledgments We would like to thank Cole Miller and the
  anonymous referee for helpful comments. SM's research was supported
by an appointment to the NASA Postdoctoral Program at the GSFC. This
material is based upon work supported by the National Aeronautics and
Space Administration under Grant No. 14-ADAP14-0198 issued through the
Science Mission Directorate.


\begin{thebibliography}{0}
\expandafter\ifx\csname natexlab\endcsname\relax\def\natexlab#1{#1}\fi

\end{thebibliography}


\begin{thebibliography}{19}

\bibitem[AlGendy 
\& Morsink(2014)]{2014ApJ...791...78A} AlGendy, M., \& Morsink, S.~M.\ 2014, \apj, 791, 78

\bibitem[Berkhout 
\& Levin(2008)]{2008MNRAS.385.1029B} Berkhout, R.~G., \& Levin, Y.\ 2008, \mnras, 385, 1029 

\bibitem[{{Bhattacharyya} \&
  {Strohmayer}(2006{\natexlab{a}})}]{2006ApJ...636L.121B}
{Bhattacharyya}, S., \& {Strohmayer}, T.~E. 2006{\natexlab{a}}, \apjl, 636,
  L121

\bibitem[{{Bhattacharyya} \&
  {Strohmayer}(2006{\natexlab{b}})}]{2006ApJ...642L.161B}
---. 2006{\natexlab{b}}, \apjl, 642, L161

\bibitem[{{Bhattacharyya} \& {Strohmayer}(2007)}]{2007ApJ...666L..85B}
---. 2007, \apjl, 666, L85

\bibitem[{{Cavecchi} {et~al.}(2013){Cavecchi}, {Watts}, {Braithwaite}, \&
  {Levin}}]{2013MNRAS.434.3526C}
{Cavecchi}, Y., {Watts}, A.~L., {Braithwaite}, J., \& {Levin}, Y. 2013, \mnras,
  434, 3526
  
\bibitem[Cavecchi et al.(2015a)]{2015MNRAS.448..445C} Cavecchi, Y., Watts, 
A.~L., Levin, Y., \& Braithwaite, J.\ 2015{\natexlab{a}}, \mnras, 448, 445 

\bibitem[Cavecchi et al.(2015b)]{2015arXiv150902497C} Cavecchi, Y., Levin, 
Y., Watts, A.~L., \& Braithwaite, J.\ 2015{\natexlab{b}}, {arXiv:1509.02497}

\bibitem[{{Chakraborty} \& {Bhattacharyya}(2014)}]{2014ApJ...792....4C}
{Chakraborty}, M., \& {Bhattacharyya}, S. 2014, \apj, 792, 4

\bibitem[{{Cooper} \& {Narayan}(2007)}]{2007ApJ...657L..29C}
{Cooper}, R.~L., \& {Narayan}, R. 2007, \apjl, 657, L29

\bibitem[{{Cumming} \& {Bildsten}(2000)}]{2000ApJ...544..453C}
{Cumming}, A., \& {Bildsten}, L. 2000, \apj, 544, 453

\bibitem[{{Galloway} {et~al.}(2008){Galloway}, {Muno}, {Hartman}, {Psaltis}, \&
  {Chakrabarty}}]{2008ApJS..179..360G}
{Galloway}, D.~K., {Muno}, M.~P., {Hartman}, J.~M., {Psaltis}, D., \&
  {Chakrabarty}, D. 2008, \apjs, 179, 360

\bibitem[{{Heyl}(2004)}]{2004ApJ...600..939H}
{Heyl}, J.~S. 2004, \apj, 600, 939

\bibitem[{{Lamb} \& {Lamb}(1978)}]{1978ApJ...220..291L}
{Lamb}, D.~Q., \& {Lamb}, F.~K. 1978, \apj, 220, 291

\bibitem[{{Lee} \& {Strohmayer}(2005)}]{2005MNRAS.361..659L}
{Lee}, U., \& {Strohmayer}, T.~E. 2005, \mnras, 361, 659

\bibitem[{{Maurer} \& {Watts}(2008)}]{2008MNRAS.383..387M}
{Maurer}, I., \& {Watts}, A.~L. 2008, \mnras, 383, 387

\bibitem[Miller \& Lamb(1998)]{1998ApJ...499L..37M} Miller, M.~C., \& Lamb, F.~K.\ 1998, \apjl, 499, L37

\bibitem[Muno et al.(2002)]{2002ApJ...580.1048M} Muno, M.~P., Chakrabarty, 
D., Galloway, D.~K., \& Psaltis, D.\ 2002, \apj, 580, 1048

\bibitem[Muno et al.(2002)]{2002ApJ...581..550M} Muno, M.~P., {\"O}zel, F., 
\& Chakrabarty, D.\ 2002, \apj, 581, 550

\bibitem[{{Piro} \& {Bildsten}(2005)}]{2005ApJ...629..438P} 
{Piro}, A.~L., \& {Bildsten}, L.\ 2005, \apj, 629, 438

\bibitem[{{Poutanen}(2012)}]{2012P}
{Poutanen}, J., private communication (LOFT Dense Matter working group)

\bibitem[{{Poutanen} \& {Beloborodov}(2006)}]{2006MNRAS.373..836P}
{Poutanen}, J., \& {Beloborodov}, A.~M. 2006, \mnras, 373, 836

\bibitem[{{Spitkovsky} {et~al.}(2002){Spitkovsky}, {Levin}, \&
  {Ushomirsky}}]{2002ApJ...566.1018S}
{Spitkovsky}, A., {Levin}, Y., \& {Ushomirsky}, G. 2002, \apj, 566, 1018

\bibitem[{{Strohmayer} \& {Bildsten}(2006)}]{2006csxs.book..113S}
{Strohmayer}, T., \& {Bildsten}, L. 2006, {New views of thermonuclear bursts},
  ed. W.~H.~G. {Lewin} \& M.~{van der Klis}, 113--156

\bibitem[{{Strohmayer} {et~al.}(1997){Strohmayer}, {Zhang}, \&
  {Swank}}]{1997ApJ...487L..77S}
{Strohmayer}, T.~E., {Zhang}, W., \& {Swank}, J.~H. 1997, \apjl, 487, L77

\bibitem[{{Strohmayer} {et~al.}(1996){Strohmayer}, {Zhang}, {Swank}, {Smale},
  {Titarchuk}, {Day}, \& {Lee}}]{1996ApJ...469L...9S}
{Strohmayer}, T.~E., {Zhang}, W., {Swank}, J.~H., {et~al.} 1996, \apjl, 469,
  L9

\bibitem[Taam(1980)]{1980ApJ...241..358T} Taam, R.~E.\ 1980, \apj, 241, 358

\bibitem[{{van Straaten} {et~al.}(2001){van Straaten}, {van der Klis},
  {Kuulkers}, \& {M{\'e}ndez}}]{2001ApJ...551..907V}
{van Straaten}, S., {van der Klis}, M., {Kuulkers}, E., \& {M{\'e}ndez}, M.
  2001, \apj, 551, 907

\bibitem[Watts(2012)]{2012ARA&A..50..609W} Watts, A.~L.\ 2012, \araa, 50, 609

\bibitem[Weinberg et al.(2006)]{2006ApJ...639.1018W} Weinberg, N.~N., 
Bildsten, L., \& Schatz, H.\ 2006, \apj, 639, 1018

\bibitem[{{Woosley} \& {Taam}(1976)}]{1976Natur.263..101W}
{Woosley}, S.~E., \& {Taam}, R.~E. 1976, \nat, 263, 101

\bibitem[Woosley et al.(2004)]{2004ApJS..151...75W} Woosley, S.~E., Heger, 
A., Cumming, A., et al.\ 2004, \apjs, 151, 75

\end{thebibliography}

\end{document}